\documentclass[12pt,preprint,notoc,nohyper]{KSJCAP} 

\usepackage{epsfig,multicol,bbm}

\newcommand\fverb{\setbox\pippobox=\hbox\bgroup\verb}
\newcommand\fverbdo{\egroup\medskip\noindent
            \fbox{\unhbox\pippobox}\ }
\newcommand\fverbit{\egroup\item[\fbox{\unhbox\pippobox}]}
\newbox\pippobox

\title{Hawking radiation of Dirac particles from black strings}

\author{Jamil Ahmed and K. Saifullah  \\

Department of Mathematics, Quaid-i-Azam University, Islamabad,
Pakistan \\

Electronic address: \email{saifullah@qau.edu.pk}}

\preprint{}  

\abstract{Hawking radiation has been studied as a phenomenon of
quantum tunneling in different black holes. In this paper we extend
this semi-classical approach to cylindrically symmetric black holes.
Using the Hamilton-Jacobi method and WKB approximation we calculate
the tunneling probabilities of incoming and outgoing Dirac particles
from the event horizon and find the Hawking temperature of these
black holes. We obtain results both for uncharged as well as charged
particles.}

\begin{document}

\section{Introduction}

The semi-classical tunneling approach to Hawking radiation which was
initially developed for the Schwarzchild black hole \cite{KF, KW,
Pa} has since then been successfully applied to a wide range of
spherically symmetric black holes or spherical black holes with an
axis of symmetry. These include the Kerr, Kerr-Newman, Taub-NUT,
G\"{o}del, dilatonic black holes and those with acceleration and
rotation \cite{KM06, KM08a, KM08b, ZL, LRW, CJZ, Ji, DJ, YY, GS,
RS}. This has proved to be a quite robust technique mathematically.
In this approach the rate of emission and absorption of particles
across the black hole horizon is calculated from the imaginary part
of their classical action \cite{SP, SPS}. This method gives the
Hawking temperature of black holes as well.

Another type of axial symmetry is cylindrical symmetry.
Cylindrically symmetric fields are axisymmetric about an infinite
axis (Killing vector, $\partial _{\theta }$) and translationally
symmetric along that axis (Killing vector, $\partial _{z}$). If we
include time translation ($\partial _{t}$) also then the stationary
cylindrically symmetric spacetimes admit three Killing vectors,
$\partial _{t}$, $\partial _{\theta }$, $\partial _{z}$ as the
minimal symmetry with the algebra $\Bbb{R} \otimes SO\left( 2\right)
\otimes \Bbb{R}$. Ever since the first investigations of
cylindrically symmetric black hole solutions
\cite{lemos95a,lemos95b, LZ, S} with negative cosmological constant
they have been a subject of interest for mathematical and physical
properties. These solutions are asymptotically anti-de Sitter in
transverse direction and along the axis. They have also been studied
in the context of supergravity theories, topological defects and low
energy string theories \cite{HT, AK, Ka} and hence the name
\emph{black strings}. These objects have also been discussed in the
presence of Born-Infeld and power Maxwell invariant fields, and in
the context of higher dimensional and Gauss-Bonnet gravity theories,
and for their thermodynamical properties \cite{BCGZ, CBSV, He, CPTZ,
FS}.

In this paper we show that the tunneling method which has thus far
been applied to spherical configurations can also be applied to
cylindrical black holes giving results that are consistent with the
literature. For this purpose we study radiation of uncharged and
charged fermions from black strings. We find the tunneling
probabilities of these particles using the WKB approximation and
then calculate Hawking temperature at the horizon.

The solution of Einsteins field equations with a negative
cosmological constant in the form of cylindrically
symmetric spacetime when mass is the only parameter is given by \cite{CZ}%
\begin{equation}
ds^{2}=-\left( \alpha ^{2}r^{2}-\frac{M}{r}\right) dt^{2}+\left(
\alpha ^{2}r^{2}-\frac{M}{r}\right) ^{-1}dr^{2}+r^{2}d\theta
^{2}+\alpha ^{2}r^{2}dz^{2},  \label{3.2}
\end{equation}
where $-\infty<t, z, <\infty$, $0\leq r <\infty$, $0\leq \theta \leq
2 \pi$, $\alpha =-\Lambda /3$, $\Lambda $ being the cosmological
constant, and $M$ is related to the ADM mass density of the black
string. We write
\begin{equation}
F\left( r\right) = \left( \alpha ^{2}r^{2}-\frac{M}{r}\right),
\label{3.13}
\end{equation}
giving the event horizon of the black hole to be
\begin{equation}
r_{+}=\left( \frac{M}{\alpha ^{2}}\right) ^{\frac{1}{3}}.
\label{3.12}
\end{equation}

\section{Tunneling of Dirac particles}

Hawking radiation from black holes comprises different types of
charged and uncharged particles. In this section we investigate
tunneling of Dirac particles from the event horizon of black string
solution via tunneling formalism. We assume that the massless spinor
field $\Psi$ obeys the general covariant Dirac's equation

\begin{equation}
\iota \gamma ^{\mu }D_{\mu }\Psi +\frac{m}{\hbar }\Psi =0,
\label{3.3}
\end{equation}
where $\hbar $ is the reduced Planck's constant, and
\[
D_{\mu }=\partial _{\mu }+\Omega _{\mu }, \,\,\, \Omega _{\mu
}=\frac{1}{2}\iota \Gamma _{\mu }^{\alpha \beta }\Sigma _{\alpha
\beta },\Sigma _{\alpha \beta }=\frac{1}{4}\iota \left[ \gamma
^{\alpha },\gamma ^{\beta }\right] .
\]%
The $\gamma ^{\mu }$ matrices satisfy $\left[ \gamma ^{\alpha
},\gamma ^{\beta }\right] =-\left[ \gamma ^{\beta},\gamma ^{\alpha
}\right]$, when $\alpha \neq \beta$ and $\left[ \gamma ^{\alpha
},\gamma ^{\beta }\right] =0$, when $\alpha =\beta$. Using this and
values of the Christoffel symbols $\Gamma _{\mu }^{\alpha \beta}$,
we note that $\Omega _{\mu }=0$, thus yielding $D_{\mu }=\partial
_{\mu }$. This reduces Dirac's equation to

\begin{equation}
\iota \gamma ^{\mu }\partial _{\mu }\Psi +\frac{m}{\hbar }\Psi =0.
\label{3.8}
\end{equation}
We can choose the $\gamma ^{\mu }$ matrices as
\begin{equation}
\gamma ^{t}=\frac{1}{\sqrt{F\left( r\right) }}\gamma ^{0}, \,\,\,
\gamma
^{r}=\sqrt{F\left( r\right) }\gamma ^{3}, \\
\gamma ^{\theta }=\frac{1}{r}\gamma ^{1}, \,\,\, \gamma
^{z}=\frac{1}{\alpha r}\gamma ^{2}, \label{3.15}
\end{equation}
where
\[
\gamma ^{0}=\left(
\begin{array}{cc}
\iota & 0 \\
0 & -\iota%
\end{array}%
\right), \,\,  \gamma ^{1}=\left(
\begin{array}{cc}
0 & \sigma ^{1} \\
\sigma ^{1} & 0%
\end{array}%
\right) ,
\]%
\begin{equation}
\gamma ^{2}=\left(
\begin{array}{cc}
0 & \sigma ^{2} \\
\sigma ^{2} & 0%
\end{array}%
 \right), \,\,   \gamma ^{3}=\left(
\begin{array}{cc}
0 & \sigma ^{3} \\
\sigma ^{3} & 0%
\end{array}%
\right). \label{3.16}
\end{equation}
Here $\sigma ^{\mu }$ are the Pauli matrices
\[
\sigma ^{1}=\left(
\begin{array}{cc}
0 & 1 \\
1 & 0%
\end{array}%
\right),  \sigma ^{2}=\left(
\begin{array}{cc}
0 & -\iota \\
\iota & 0%
\end{array}%
\right), \sigma ^{3}=\left(
\begin{array}{cc}
1 & 0 \\
0 & -1%
\end{array}%
\right) .
\]%
In order to find the solution of Eq. (\ref{3.8}) in the background
of the black string (Eq. (\ref{3.2})) we employ the following ansatz
for Dirac's field, corresponding to the spin up case
\begin{equation}
\Psi _{\uparrow }\left( t,r,\theta ,z\right) =\left(
\begin{array}{c}
A\left( t,r,\theta ,z\right) \xi _{\uparrow } \\
B\left( t,r,\theta ,z\right) \xi _{\uparrow }%
\end{array}%
\right) \exp \left( \frac{\iota }{\hbar }I\right) ,  \label{3.17}
\end{equation}
where $\xi _{\uparrow }=\left(
\begin{array}{c}
1 \\
0
\end{array}
\right)$, $I$ represents the classical action, and $A$ and $B$ are
arbitrary functions of the coordinates. Using the above values we
evaluate the four terms of Eq. (\ref{3.8}) one by one and apply WKB
approximation. We divide by the exponential term and multiply by
$\hbar $. Considering terms only upto the leading order in $\hbar $,
this procedure yields the following set of four equations.
\begin{equation}
 \frac{- \iota A I_{t}}{\sqrt{F\left( r\right) }}- B \sqrt{F\left(
r\right) } I_{r} +mA=0,   \label{3.24}
\end{equation}
\begin{equation}
-B\left( \frac{I_{\theta }}{r}+\frac{\iota I_{z}}{\alpha r}\right)
=0, \label{3.25}
\end{equation}
\begin{equation}
 \frac{\iota B I_{t}}{\sqrt{F\left( r\right) }}-A \sqrt{F\left(
r\right) } I_{r} +mB=0,   \label{3.26}
\end{equation}
\begin{equation}
-A\left( \frac{I_{\theta }}{r}+\frac{\iota I_{z}}{\alpha r}\right)
=0. \label{3.27}
\end{equation}
The derivatives involving higher order in $\hbar $ have been
neglected as we are taking only the lowest order in WKB
approximation. Taking into account the Killing vectors of the
background spacetime we can employ the following ansatz for the
action in the spin up case
\begin{equation}
I_{\uparrow }\left( t,r,\theta ,z\right) =-Et+l\theta +Jz+W\left(
r\right),    \label{3.28}
\end{equation}
where $E$ is the energy of the emitted particles and $W$ is the part
of the action $I_{\uparrow }$ that contributes to the tunneling
probability. Using this ansatz in Eqs. (\ref{3.24})$-$(\ref{3.27})
and noting that the contribution of $J$ and $l$ to the imaginary
part of the action is canceled out we neglect the equations
containing $J$ and $l$. So, for the massless case, $m=0$, the
function $W(r)$ can be calculated only from the following equations
\begin{equation}
 -\iota A E+B F(r) \frac{dW}{dr} =0, \label{3.33}
\end{equation}
\begin{equation}
\iota B E+A F(r)\frac{dW}{dr} =0. \label{3.34}
\end{equation}
In this case we get $A=-\iota B$ and

\begin{equation}
\left( \frac{dW_{+}}{dr}\right) =\frac{E}{F\left( r\right) }.
\label{3.37}
\end{equation}
Similarly, the other equation yields $A=\iota B$ and
\begin{equation}
\left( \frac{dW_{-}}{dr}\right) =\frac{-E}{F\left( r\right) }.
\label{3.38}
\end{equation}
%
%
In order to integrate Eq. (\ref{3.37}) we write it as

\begin{equation}
W_{+}\left( r\right) =\int \frac{E}{F(r)}dr.
\end{equation}
In this integral $r=r_{+}$ is a simple pole, therefore integrating
around the pole we get

\begin{equation}
W_{+}\left( r\right) =\frac{\pi \iota E}{2\alpha
^{2}r_{+}+\frac{M}{r_{+}^{2} }}. \label{3.43}
\end{equation}
Similarly
\begin{equation}
W_{-}\left( r\right) =\frac{-\pi \iota E}{2\alpha
^{2}r_{+}+\frac{M}{r_{+}^{2} }}. \label{3.44}
\end{equation}
The probabilities of crossing the horizon in each direction can be
given by \cite{SP, SPS}
\[
P_{emission} \varpropto \exp \left( -2Im  I\right) =\exp \left( -2Im
W_{+}\right),
\]
\[
P_{absorption} \varpropto \exp \left( -2Im  I\right) =\exp \left(
-2Im  W_{-}\right) .
\]%
While computing the imaginary part of the action, we note that it is
same for both the incoming and outgoing solutions. Now the
probability of particles tunneling from inside to outside the
horizon is given by
\[
\Gamma \varpropto \frac{P_{emission} }{P_{absorption} }=\frac{ \exp
\left( -2 Im  W_{+}\right) }{\exp \left( -2 Im W_{-}\right) },
\]
which gives
\begin{equation}
\Gamma =\exp \left( -4Im  W_{+}\right) . \label{3.45}
\end{equation}
Using the value of $W_{+}$ this becomes

\begin{equation}
\Gamma =\exp \left( \frac{-4\pi E}{2\alpha ^{2}r_{+}+\frac{M}{r_{+}^{2}}}%
\right) . \label{3.46}
\end{equation}
Now we know that the tunneling probability is given by $\Gamma =\exp
\left( -\beta E\right)$, where $\beta =\frac{1}{T_{H}}$, yielding

\begin{equation}
T_{H}=\frac{1}{4\pi }\left( 2\alpha
^{2}r_{+}+\frac{M}{r_{+}^{2}}\right) , \label{3.48}
\end{equation}
which is the correct Hawking temperature for black strings \cite{FS,
CZ}. For the massive case (i.e. $m\neq 0$) we get

\[
\left(\frac{A}{B}\right)^2=\frac{-\iota E+m \sqrt{F\left( r\right)
}}{\iota E+m \sqrt{F\left( r\right) }}.
\]%
Near the horizon we get $A^2=-B^2$, and obtain the same Hawking
temperature as in the massless case. This is because near the
horizon the massive particles behave like massless particles.

\section{Tunneling from the charged black string}

Soon after the discovery of the cylindrically symmetric black hole
solution it was extended to the case with electromagnetic field
\cite{CZ}. Here we study quantum tunneling of Dirac particles from
charged black strings. The line element can be written as
\begin{eqnarray}
ds^{2} &=&-\left( \alpha ^{2}r^{2}-\frac{4M}{\alpha
r}+\frac{4Q^{2}}{\alpha
^{2}r^{2}}\right) dt^{2}+\left( \alpha ^{2}r^{2}-\frac{4M}{\alpha r}+\frac{%
4Q^{2}}{\alpha ^{2}r^{2}}\right) ^{-1}dr^{2}  \nonumber \\
&&+r^{2}d\theta ^{2}+\alpha ^{2}r^{2}dz^{2},  \label{4.1}
\end{eqnarray}%
with the scalar potential%
\begin{equation}
F_{tr}=\frac{-2Q}{\alpha r}. \label{4.2}
\end{equation}
Here $M$ is mass, $Q$ is charge, $\alpha =-\Lambda /3$. Putting
$g^{11}=0$ gives the location of the inner and outer horizons
\cite{ZY} for the non-extremal charged black string as

\begin{equation}
r_{\pm }=\frac{1}{2}\left[ \sqrt{2R}\pm (-2R+\frac{8M}{\alpha ^{3}\sqrt{2R}}%
)^{\frac{1}{2}}\right],  \label{4.3}
\end{equation}
where
\begin{equation}
R=\left\{ \frac{M^{2}}{\alpha ^{6}}+\left[ (\frac{M^{2}}{\alpha
^{6}})^{2}-( \frac{4Q^{2}}{3\alpha ^{4}})^{3}\right]
^{\frac{1}{2}}\right\} ^{\frac{1}{3} }+\left\{ \frac{M^{2}}{\alpha
^{6}}-\left[ (\frac{M^{2}}{\alpha ^{6}})^{2}-( \frac{4Q^{2}}{3\alpha
^{4}})^{3}\right] ^{\frac{1}{2}}\right\} ^{\frac{1}{3}}.
   \label{4.4}
\end{equation}
In this case the function $F(r)$ takes the following form
\begin{eqnarray}
F(r)=\frac{\alpha ^{4}r^{4}-4M\alpha r+4Q^{2}}{\alpha ^{2}r^{2}}
\label{f},    \label{g}
\end{eqnarray}
The charged Dirac equation for a particle with charge $q$ is given
by
\begin{equation}
\iota \gamma ^{\mu }\left( D_{\mu }-\frac{\iota q}{\hbar }A_{\mu
}\right) \Psi +\frac{m}{\hbar }\Psi =0.   \label{4.5}
\end{equation}
We choose the $\gamma ^{\mu }$ matrices as before but noting that
the function $F$ is now given by Eq. (\ref{f}). Assuming the Dirac
field of the form taken earlier and applying WKB approximation we
obtain the following set of four equations
\begin{equation}
- \iota A \frac{I_{t}}{\sqrt{F\left( r\right) }}-B \sqrt{F\left(
r\right) } I_{r}-2\iota A \frac{Qq}{\alpha r\sqrt{F\left( r\right)
}} +mA=0, \label{4.15}
\end{equation}
\begin{equation}
B\left( \frac{I_{\theta}}{r}+\frac{\iota I_{z}}{\alpha r}\right) =0,
\label{4.16}
\end{equation}
\begin{equation}
\iota B  \frac{I_{t}}{\sqrt{F\left( r\right) }}-A \sqrt{F\left(
r\right) } I_{r}+2 \iota B \frac{Qq}{\alpha r\sqrt{F\left( r\right)
}} +mB=0, \label{4.17}
\end{equation}
\begin{equation}
A\left( \frac{I_{\theta}}{r}+\frac{\iota I_{z}}{\alpha r}\right) =0.
\label{4.18}
\end{equation}
Substituting the same form of the action as earlier and following an
analysis similar the one done in the uncharged case we find that

\begin{equation}
A=-\iota B \,\,\,\,\,\, \textrm{and} \,\,\,\,\,\,
\frac{dW}{dr}=\frac{E\alpha r-2Qq }{\alpha r F\left( r\right) }.
\label{400}
\end{equation}
Using the value of $F(r)$ in Eq. (\ref{400}) and integrating by
employing the residue theory we get

\begin{equation}
W_{+}=\pi \iota \left( \frac{E\alpha ^{2}r_{+}^{3}-2Q\alpha r_{+}^{2}q}{%
2\alpha ^{4}r_{+}^{4}+4M\alpha r_{+}-8Q^{2}}\right) . \label{4.28}
\end{equation}
Also
\[
W_{-}=\pi \iota \left( \frac{-E\alpha ^{2}r_{+}^{3}+2Q\alpha r_{+}^{2}q}{%
2\alpha ^{4}r_{+}^{4}+4M\alpha r_{+}-8Q^{2}}\right) .
\]
Now, as before we write the probability of particles tunneling from
inside to outside the horizon as
\begin{equation}
\Gamma = \exp \left[ \frac{-4\pi\left( E\alpha ^{2}r_{+}^{3}-2Q\alpha r_{+}^{2}q\right)}{%
2\alpha ^{4}r_{+}^{4}+4M\alpha r_{+}-8Q^{2}}\right].  \label{4.31}
\end{equation}
Comparing this with $\Gamma = \exp (-E/T_H)$, the Hawking
temperature for charged black string takes the form

\begin{equation}
T_{H}=\frac{1}{2\pi }\left( \alpha ^{2}r_{+}+\frac{2M}{\alpha r_{+}^{2}}-%
\frac{4Q^{2}}{\alpha^2 r_{+}^{3}}\right) . \label{4.33}
\end{equation}
which is consistent with the literature \cite{FS, CZ}.

\section{Conclusion}

In the tunneling formalism the probability of particles crossing the
black hole horizon on either sides are calculated using complex path
integrals. The particles traverse geodesics which are forbidden in
classical treatments. However the probability for absorption of
particles should actually be equal to 1 as this is a path which is
permitted classically \cite{KF}. This provides an efficient way of
computing the Hawking temperature as well. This technique has been
successfully used for spherically symmetric black holes. We have
extended its application to cylindrically symmetric configurations.
Solving Dirac's equations in the background of uncharged and charged
black strings and applying WKB approximation, we have calculated the
tunneling probability of fermions. Hawking temperature in both the
cases is also calculated.


\end{document}